\documentstyle[aps,preprint,floats,psfig]{revtex}
\begin{document}
\newcommand {\be}{\begin{equation}}
\newcommand {\ee}{\end{equation}}
\newcommand {\bea}{\begin{eqnarray}}
\newcommand {\eea}{\end{eqnarray}}
\newcommand {\nn}{\nonumber}

\draft
%
%
%
%

\title{Anisotropic s-wave superconductivity in MgB$_2$
}

\author{Stephan Haas and Kazumi Maki}
\address{Department of Physics and Astronomy, University of Southern
California, Los Angeles, CA 90089-0484
}

\date{\today}
\maketitle

\begin{abstract}
  
It has recently been observed that MgB$_2$ is a 
superconductor with a high transition temperature.
Here we propose a model of anisotropic s-wave superconductivity 
which consistently describes the observed properties of this compound, 
including the thermodynamic and optical response 
in sintered MgB$_2$ wires. 
We also determine the shape of the quasiparticle density of states and
the anisotropy of the upper critical field and the superfluid
density which should be detectable once single-crystal samples
become available. 

\end{abstract}

\noindent{\it 1. Introduction}

The recently discovered superconductor MgB$_2$ has received
great attention, mainly because of its high
transition temperature of approximately 39K.
\cite{akimitsu,budko,walti,wang,gorshunov,budko2}
This relatively high $\rm T_c$ value may not be
too surprising since MgB$_2$ involves some of the lightest 
elements in the periodic table, thus implying a large
Debye phonon frequency for this compound. Indeed, data from 
thermodynamic measurements\cite{budko,walti}
suggest a Debye frequency $\Theta_D \approx$ 750 K
as well as a significant isotope effect with $\alpha = 0.26$. 

More surprisingly, a rather small zero-temperature
energy gap $\Delta_0 = 1.20 k_B T_c$
was deduced from  measurements of the specific heat\cite{walti,wang}
and from the optical conductivity\cite{gorshunov}. Presently, we do not 
know of any other examples for such a small ratio $\Delta_0/k_B T_c $
in conventional s-wave superconductors. This observation  
naturally suggests an anisotropic s-wave order parameter for this 
material, where the energy gap detected in the thermodynamic measurements
is in fact the minimum of the superconducting gap function. 
An alternative explanation has been suggested in terms of 
multi-band models.\cite{bascones,shulga} In many respects the proposed
phenomenological model with an anisotropic energy gap is similar  
to a multi-band model. In fact, the latter may be thought 
of as a discretized version of the former.
More recently, the upper critical critical field for a dense
wire of MgB$_2$ was inferred from the field dependence of the 
critical current.\cite{budko2} From these measurements it was 
concluded that MgB$_2$ is an extreme type II superconductor
with a Ginzburg-Landau parameter $\kappa \simeq 23$.

Drafting an effective model with an anisotropic s-wave order parameter 
for MgB$_2$, the first important question to address is the 
direction and the magnitude
of the anisotropy over the approximately ellipsoidal 
Fermi surface. In the distantly related tetragonal systems $\rm YNi_2B_2C$
and $\rm LuNi_2B_2C$, such an anisotropy 
has been shown to appear in the a-b planes.\cite{canfield,wang2}
On the other hand, MgB$_2$ has a hexagonal crystal structure, 
and thus the anisotropy is most likely to occur along the 
c-direction similar to other hexagonal crystals, such as $\rm UPt_3$
\cite{heffner} and $\rm UPd_2Al_3$\cite{jourdan}, and tetragonal
crystals, such as $\rm Sr_2RuO_4$\cite{won,izawa}.   

We therefore propose a BCS model for MgB$_2$ with a superconducting 
order parameter given by
\bea
\Delta({\bf k}) = \Delta \left( \frac{1 + a z^2}{1 + a} \right).
\eea
where the parameter $a$ determines the anisotropy, $z = \cos{\theta}$,
and $\theta$ is the polar angle. For the calculations
outlined in this paper, we impose the experimentally observed gap ratio,
\bea
\frac{\Delta_{min}(T = 0)}{k_B T_c } = \frac{\Delta_0}{(1 + a ) k_B T_c }
= 1.20,
\eea
which in turn yields $a \simeq 1$. In the following, we will therefore explore 
the consequences of anisotropic s-wave superconductivity in MgB$_2$
within the framework of weak-coupling BCS theory, setting $a =1$. 

\noindent{\it 2. Density of States}

In Fig. 1(a) the anisotropic s-wave order parameter
$\Delta({\bf k}) = \Delta (1 + z^2)/2$, is plotted in momentum space. 
This function is an ellipsoid with a minor axis of length
$\Delta/2$ within the 
a-b plane, and a major axis of length $\Delta$ along the c-direction. 
The smaller magnitude of the gap function within the a-b plane is 
consistent with the stronger in-plane Coulomb repulsion that has been 
suggested in Ref. \cite{voelker}.  

\begin{figure}[h]
\centerline{\psfig{figure=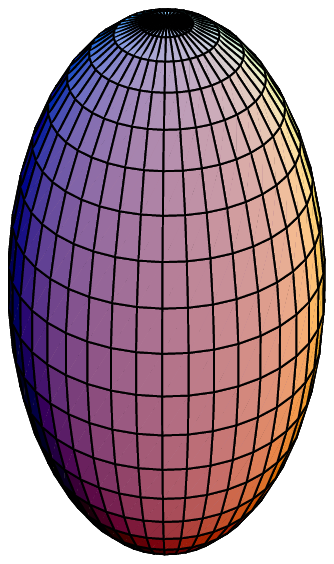,width=5cm,angle=0}
\psfig{figure=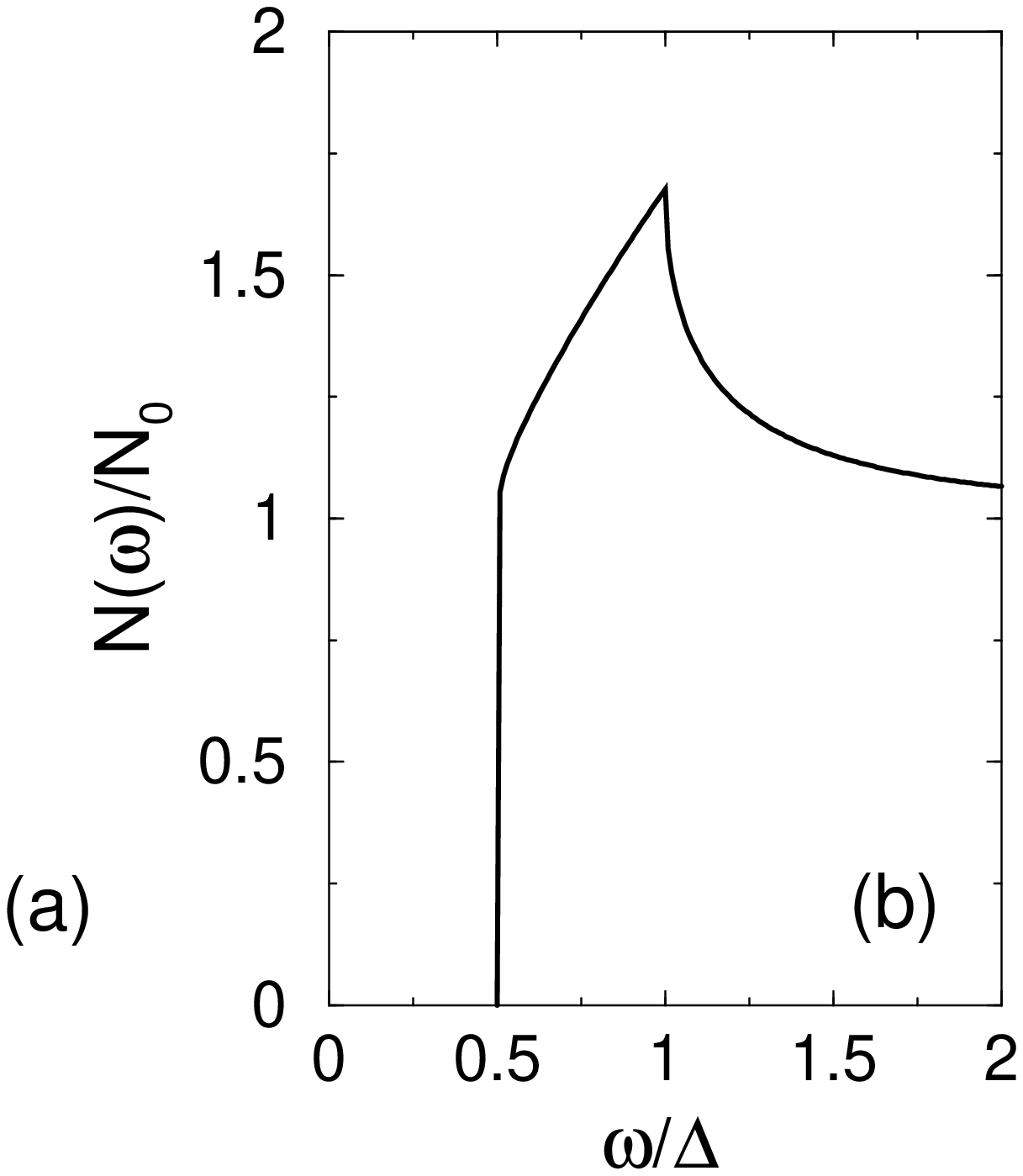,width=6cm,angle=0}}
\vspace{0.3cm}
\caption{
(a) Anisotropic s-wave order parameter. 
(b) Quasiparticle 
density of states.
}
\end{figure}

The corresponding density of states can be calculated from $\Delta({\bf k})$ 
within weak-coupling theory. It is given by
\bea
\frac{N(E)}{N_0} & = & \int_0^1 dz \Re e \left( \frac{|E|} 
{\sqrt{E^2 - \Delta^2 (1 + z^2)^2/4}} \right) \nn \\
 & = & 0 \ \ \ \ \ \ \ \ \ \ \ \ \ \ \ \ \ \ \ \ \ \ \ \ \ \ \ \ \ \ \ \ \ \ \ 
 \ \ \ \ \ \ \ \ \ \ \ \ \ \ \ \ \ \ \ \ \ \ \ \ \ \  
{\rm if} \ \  0 < E < \Delta/2, \nn \\
 & = & \sqrt{\frac{E}{\Delta}} K\left(\sqrt{\frac{2 E - \Delta}{4 E}}\right)
 \ \ \ \ \ \ \ \ \ \ \ \ \ \ \ \ \ \ \ \ \ \ \ \ \ \ \ \ \ \ \ \ \ \ \ 
{\rm if} \ \  \Delta/2 < E < \Delta, \\
& = & \sqrt{\frac{E}{\Delta}} F\left(\sin^{-1}\sqrt{\frac{2 E \Delta}
{(2E - \Delta)(E + \Delta)}} , \sqrt{\frac{2 E - \Delta}{4 E}} \right)
 \ \ \ \ \ \ 
{\rm if} \ \  E > \Delta,  \nn  
\eea
where $K(k)$ and $F(\Phi , k)$ are the complete and incomplete
elliptic integrals of the first kind. This quasiparticle density of states  
is shown in Fig. 1(b). It is fully gapped with an onset of spectral
weight at the minimum gap value $E = \Delta/2$
and a cusp at $E = \Delta$.

\noindent{\it 3. Thermodynamics}

\begin{figure}[h]
\centerline{\psfig{figure=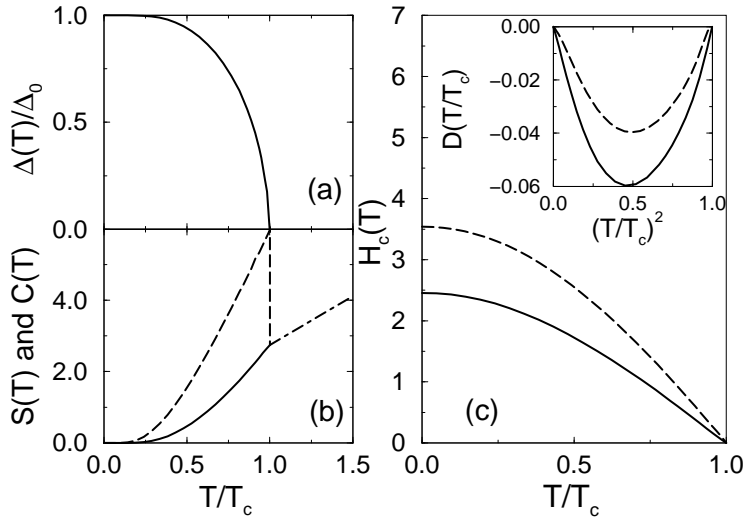,width=9cm,angle=0}}
\vspace{0.6cm}
\caption{
(a) Energy gap in an
anisotropic s-wave superconductor.
(b) Entropy (solid line) and specific
heat (dashed line). Above $\rm T_c$ these quantities
are equal (dot-dashed line).
(c) Thermodynamic critical field for anisotropic
(solid line) and isotropic (dashed line) s-wave superconductors.
Inset: deviation of these critical fields from a
parabolic temperature dependence.
}
\end{figure}

Following the formalism by Bardeen, Cooper, and Schrieffer
\cite{bardeen}, we calculate the temperature dependence of the
energy gap $\Delta (T)$, the entropy $S(T)$, the 
specific heat $C(T)$, and the thermodynamic critical field $H_c(T)$ for the 
anisotropic s-wave superconductor. Here, the entropy is given by
\bea
S(T)  = -4 \int_0^{\infty} dE N(E) \left( f \ln{f} + (1 - f) \ln{(1 - f)}
\right), \\
\eea
where $ f \equiv (1 + \exp{\beta E} )^{-1}$. Furthermore, the specific heat 
and the upper critical field are obtained from
\bea
C(T)  =  T \frac{\partial S(T)}{\partial T} \ \ \ \ \ \ \ \ {\rm and}
\ \ \ \ \ \  
\frac{H^2_c(T)}{8 \pi}   =   \int_T^{T_c} dT S(T). 
\eea
These results are displayed in Fig. 2.
As expected, the characteristic jump of the specific heat at $\rm T_c$,
$\Delta C / C_N = 1.18$, 
is small compared with the value
1.43 for an isotropic s-wave superconductor. This is
consistent with the experimental picture for MgB$_2$.\cite{walti,wang}
Furthermore, it is seen in Fig. 2(b) 
that the thermodynamic critical field is
also reduced in the anisotropic case. The deviation of the 
critical field from a parabolic temperature dependence, defined by
\bea
D(T/T_c) \equiv \frac{H_c(T)}{H_c(0)} - \left[1 - 
\left( \frac{T}{T_c} \right)^2 \right],
\eea
is shown in the inset of Fig. 2(c). The magnitude of this deviation is
substantially larger than for the isotropic case, again
consistent with the experiments.\cite{wang}  

\begin{figure}[h]
\centerline{\psfig{figure=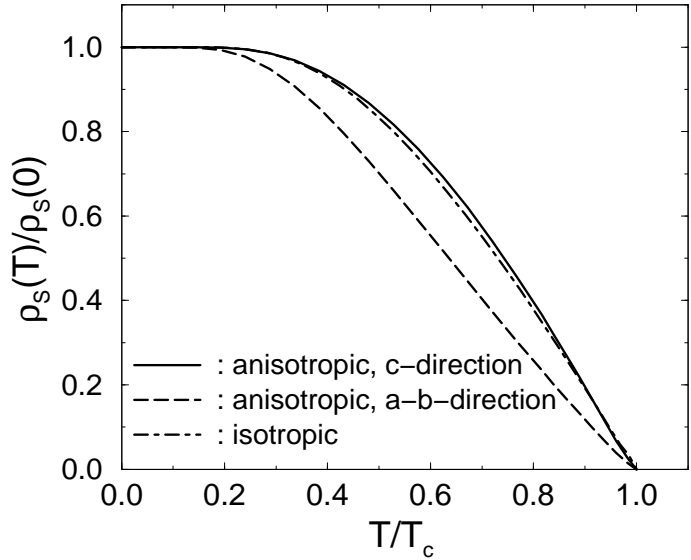,width=9cm,angle=0}}
\caption{
Superfluid densities of anisotropic and of isotropic s-wave
superconductors. 
}
\end{figure}

Finally, the superfluid density $\rho_s(T)$ shown in Fig. 3 is found to 
be rather anisotropic. Similar to the energy gap and the 
thermodynamic critical field, upon increasing the temperature
its departure from its zero-temperature value is exponentially
small in the low-temperature regime. On the other hand, in the vicinity
of $\rm T_c$ the superfluid density for the isotropic
case is linear in temperature, whereas for the anisotropic case
$\rho_s (T)$ is observed to be
slightly concave along the c-direction and slightly 
convex in the a-b plane.  
Once single crystals of MgB$_2$ become available, this anisotropy 
in the superfluid density should serve as a signature for 
the order parameter symmetry.

\noindent{\it 4. Upper Critical Field}

Recent measurements of the upper critical field
in sintered samples of MgB$_2$ have reported
a somewhat unusual temperature dependence.\cite{budko2,shulga} 
This is likely
related to the apparent anisotropy of the superconducting order 
parameter, and should thus become even more evident in measurements of the 
upper critical field in single crystals. Therefore we consider here
the two cases of an external magnetic field parallel and perpendicular 
to the crystal c-axis of MgB$_2$.

\begin{figure}[h]
\centerline{\psfig{figure=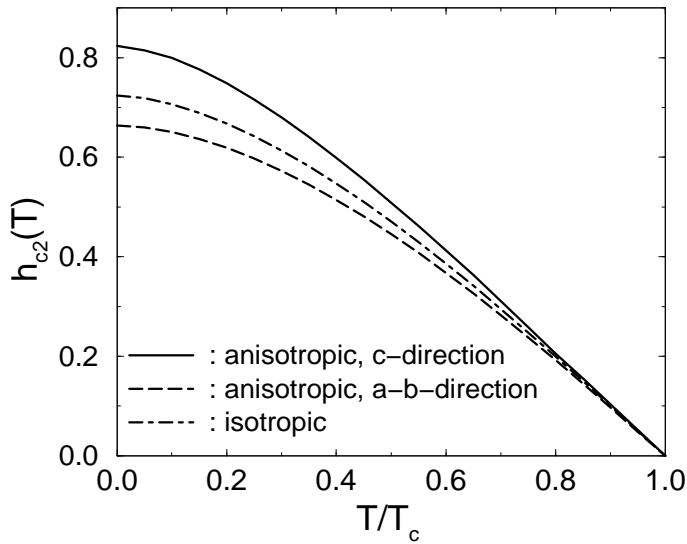,width=9cm,angle=0}}
\caption{
Upper critical fields of anisotropic and isotropic s-wave
superconductors. 
}
\end{figure}

\noindent{\it (a) ${\bf H} \parallel {\bf c}$} 

The upper critical field along the crystal c-direction
is determined from
\cite{wang2,won2},
\bea
- \ln{\frac{T}{T_c}} = \int_0^{\infty} \frac{d u}{\sinh{u}}
 \left( 1 - \frac{15}{28}
\int_0^1 dz (1 + z^2)^2 \exp{(-\rho u^2 (1 - z^2))} \right),
\eea
where $\rho \equiv (v_F^2 e H_{c2}(T))/(2(2\pi T)^2)$ and $z \equiv
\cos{\theta}$. In Fig. 4, the solution of this integral equation
normalized by its derivative at $T_c$,
$h_{c2}(T) \equiv H_{c2}(T)/(-\partial H_{c2}(T)/\partial T)|_{T_c}$, 
is plotted along with the $h_{c2} (T)$-curves
for the a-b-direction and for the isotropic case.
In the limit $T \rightarrow 0$ we obtain $h^c_{c2}(0) \simeq 0.824$ which is 
somewhat larger than the corresponding value for the isotropic s-wave 
superconductor, $h_{c2}(0) \simeq 0.728$. In the opposite limit, 
$T \rightarrow T_c$, $h^c_{c2}(T)$ exhibits a rather linear temperature
dependence within the weak-coupling BCS model. This behavior is also 
observed for the isotropic s-wave superconductor.

\noindent{\it (b) ${\bf H} \parallel {\bf a}$}

For the upper critical field within the a-b plane 
a mixing occurs of the zeroth and the second Landau levels,
leading to
\cite{wang2,won2}
\bea
\Delta ({\bf k},{\bf r}) \sim \Delta \left(
1 + C (a^{\dagger})^2 \right) | 0 \rangle , 
\eea
where $| 0 \rangle$ is the Abrikosov state, $a^{\dagger}$ is 
the raising operator, and $C$ is the mixing coefficient between the 
Landau levels which has to be determined self-consistently along
with the critical field. This leads to a set of 
coupled integral equations,
\bea 
- \ln{\frac{T}{T_c}} & = & \int_0^{\infty} \frac{d u}{\sinh{u}}
\left( 1 - \frac{15}{28}
\int_0^1 dz 
\left[ (1 + \sin^2{\theta} + \frac{3}{8} \sin^4{\theta} ) \right. \right.\nn \\
& + & \left. \left.
 \frac{1}{2} C \rho u^2 \sin^4{\theta} (1 + \frac{1}{2} \sin^2{\theta})
\right]  
\exp{(-\rho u^2 (1 - z^2))} \right),\\
- C \ln{\frac{T}{T_c}} & = & \int_0^{\infty} \frac{d u}{\sinh{u}}
\left( C - \frac{15}{28}
\int_0^1 dz
\left[ \frac{1}{4} \rho u^2 \sin^4{\theta} (1 + \frac{1}{2} \sin^2{\theta})
\right. \right.\nn \\
& + & \left. \left.
C (1 + \sin^2{\theta} + \frac{3}{8} \sin^4{\theta} )
(1 - 4 \rho u^2 \sin^2{\theta} + 2 \rho^2 u^4 \sin^4{\theta} ) \right]
\exp{(-\rho u^2 (1 - z^2))} \right),
\eea
where again $\rho \equiv (v_F^2 e H_{c2}(T))/(2(2\pi T)^2)$, and 
$\sin^2{\theta} = 1 - z^2$.
The numerical solution of this set of equations is shown in Fig. 4. The 
mixing coefficient (not shown) is found to decrease monotonously from
0.069 down to 0.062 as the temperature is increased. This implies a 
relevant admixture of the second Landau level. In the zero-temperature
limit, we obtain $h^a_{c2}(0) \simeq 0.664$. By comparing $h^c_{c2}(0)$
with $h^a_{c2}(0)$, we can thus conclude from the model calculation
that the critical field 
observed in sintered MgB$_2$ represents $H^c_{c2}(T)$ because
$H^c_{c2}(T) > H^a_{c2}(T)$ for all temperatures. 
We also note that $H^c_{c2}(T), H^a_{c2}(T) \gg H_{c}(T)$ since
for sintered MgB$_2$ wires the Ginzburg-Landau parameter was observed
to be very large $\kappa \simeq 23$.\cite{budko2}
  
\noindent{\it 5. Conclusions} 

Based on presently available experimental data on MgB$_2$ we have 
constructed a model of anisotropic s-wave superconductivity for 
this compound with $\Delta_{min}/\Delta_{max}$ = 1/2.
This simple theory appears to account rather well 
for the thermodynamic properties\cite{walti,wang}, 
the optical measurements,\cite{gorshunov} and 
for the upper critical field data\cite{budko,budko2,shulga}
on sintered samples of MgB$_2$.
Experiments on single crystal samples will clearly be important
to further investigate the applicability of this model.  

We thank M. R. Norman, B. Normand, H. Won, 
and N. C. Yeh for useful discussions, 
and acknowledge financial support by the Office of Naval Research,
Grant No. N000140110277, and by the National Science Foundation,
Grant No. DMR-0089882.

\end{document}